\documentclass[twocolumn,showpacs]{revtex4}
\usepackage{latexsym}
\usepackage{graphics}
\usepackage{epsf}
\usepackage{epsfig}
\usepackage{graphicx}
\usepackage{dcolumn}
\usepackage{bm}
\usepackage{latexsym}
\usepackage{booktabs}
\usepackage{amssymb}

\newcommand {\be}{\begin{equation}}
\newcommand {\ee}{\end{equation}}
\newcommand {\bea}{\begin{eqnarray}}
\newcommand {\eea}{\end{eqnarray}}

\newcommand {\nonum}{\nonumber}
\newcommand {\PRE}[1]{{\it Phys. Rev. E} {\bf {#1}}}
\newcommand {\PRL}[1]{{\it Phys. Rev. Lett.} {\bf {#1}}}

\begin{document}


\title{Synchronization of R{\"o}ssler Oscillators on Scale-free Topologies}

\author{Soon-Hyung Yook and Hildegard Meyer-Ortmanns}
\affiliation{School of Engineering and Science,
International University Bremen,\\
P.O.Box 750561, D-28725 Bremen, Germany}
\author{}
\affiliation{}

\date{\today}

\begin{abstract}
  We study the synchronization of R{\"o}ssler
oscillators as prototype of chaotic systems, when they are coupled
on scale-free complex networks. We find that the underlying
topology crucially affects the global synchronization properties.
Especially, we show that the existence of loops facilitates the
synchronizability of the system, whereas R\"ossler oscillators do
not synchronize on tree-like topologies beyond a certain size. By
considering Cayley trees, modified by various shortcuts, we find
that also the distribution of shortest path lengths between two
oscillators plays an important role for the global
synchronization.
\end{abstract}


\maketitle
\section{Introduction}
Synchronization is an ubiquitous phenomenon in nature, ranging from flashing fireflies in the Australian
forest \cite{fireflies}, crickets
chirping in unison \cite{crickets} in natural systems, tremor in Parkinson's disease or epilepsy in medical applications \cite{parkinson},
laser arrays \cite{laser}, or Josephson junctions in physics \cite{josephson}, electrochemical oscillators in
chemistry \cite{electrochemical} and designed synchronization in robotics.
In particular, synchronization properties of limit-cycle oscillators were studied in a number of
papers (for a review see \cite{kuramotoreview}), but even systems which are individually chaotic like R\"ossler oscillators,
can synchronize under certain conditions. R\"ossler oscillators are sometimes treated as prototype of chaotic systems.
According to a conjecture of Calenbuhr and Mikhailov \cite{mikhailovbook}, the behavior of R\"ossler oscillators shows some
universal features. For a certain class
of interactions
and under the influence of noise, clusters of synchronizing oscillators form above a certain threshold of the coupling
strength between the oscillators, and, for larger couplings, after an intermittent phase, the whole set of oscillators
synchronizes.

R\"ossler oscillators were studied for different interaction schemes and on different geometries \cite{roesslerother}.
In this paper
we study the conditions for R\"ossler oscillators to synchronize on scale-free network topologies. Scale-free networks
seem to be realized in a number of natural and artificial systems like genetic or proteomic networks, the world-wide-web
and the internet. Synchronization is certainly one of the important dynamical processes, running on these networks,
as it is supposed to be a necessary ingredient for the efficient organization and functioning of coupled individual units,
that, after all, lead to well coordinated behavior in time. Therefore we are interested in the compatibility of
scale-free topologies with synchronization, in particular for the case that the individual dynamics is chaotic.
While usually the synchronization transition is studied as a function of the coupling strength or the system size,
we describe here, in addition to the usual approach, a transition to the synchronized phase as a function of the topology.
As we shall see, when a tree becomes too large in size to allow for synchronization, synchronization  becomes possible again
beyond a critical threshold in the coupling, when a critical number of loops and shortcuts is introduced into the tree,
while it is impossible for arbitrary couplings on a tree beyond a certain size.

In section I we introduce the model and define
the order parameters that are used to distinguish the phases with and without the condensates of synchronized oscillators.
In the second section we describe details of the simulations and summarize the results in section III.

\section{The Model}

We consider a system of $N$ R\"ossler oscillators, distributed on the nodes of a scale-free network, generated with the
growth algorithm of B\'arabasi and Albert\cite{albert02} (see below). Each individual oscillator is described by the following
set of dynamical equations that was originally proposed by R\"ossler \cite{Roessler76} as a "model of a model" for describing
the trajectory of flow, satisfying
the Lorenz equation \cite{Lorenz},
\bea\label{1}
  \nonum
  \dot x&=&-\omega y - z\\
  \nonum
  \dot y &=& \omega x\;+\;a y\\
  \label{Zanette}
  \dot z&=& b- c\;z + x\;z\;.
\eea
For $\omega = 1$, $a=0.15$, $b=0.2$, $c=8.5$ the system is in the chaotic state. Among various possibilities of
coupling these oscillators, we choose
\bea\label{2}
  \nonum
  \dot x_i&=&- y_i\; -\; z_i\\
  \nonum
  \dot y_i&=& x_i\;+\;a y_i\;+\;\epsilon\;(\bar y_i -y_i)\\
  \dot z_i&=&b+(x_i-c)z_i+\epsilon(\bar x_i \bar z_i
    -x_i z_i)\;,
\eea
where $\bar x_i $, $\bar y_i $, $\bar z_i $ are averages defined as
\be\label{3}
\bar x_i \;=\;\frac{1}{k_i}\;\sum_{j=0}^{N}\;A_{ij}\;x_j
\ee
and accordingly for $\bar y_i$ and $\bar z_i$, $k_i$ denotes the degree of node $i$, $A_{ij}$ is the adjacency matrix
i.e.$A_{ij}=1$ if $i$ and $j$ are connected and $0$ otherwise, that is the only place where the topology of the network enters.
For $k_i = N-1$ and $A_{ij} = 1$
for all $i,j \in 1,...,N$ the system corresponds to a globally coupled population of R\"ossler oscillators as it
was considered in \cite{zanette98}.
In our description the
population is partially coupled rather than globally. It is coupled along
the links of the scale-free network, therefore
the driving force towards the common synchronized state
is produced by nearest-neighbors, whose number varies according to the scale-free degree-distribution.
Since our averages are still node-dependent,
the stability analysis of \cite{zanette98}, derived for $\bar x = \frac{1}{N} \sum_{j = 0}^{N} x_j$, ($\bar y, \bar z$
alike,) does not immediately apply. For this case of global coupling with driving force that tries to reduce the
difference from the common synchronized state $(\bar x, \bar y, \bar z)$, one expects a globally synchronized
stable state for $\epsilon = 1, a<1$, so that all deviations from global averages exponentially decrease with time \cite{zanette98}.
In our scheme the force drives to node-dependent average values over nearest neighbors whose number is neither regular
nor $N-1$, i.e. all-to-all. Nevertheless we find a result quite similar to the all-to-all case: a global attractor
to a synchronized state as long as $\epsilon < 1.25$. The stability is evident on the level of numerical simulations.

In order to check how the results depend on the non-linear terms of our coupling scheme, we made also some tests for
the linear vector coupling defined according to
\bea\label{4}
  \nonum
  \dot x_i&=&-\;y_i - z_i\;+\;\epsilon\;(\bar x_i - x_i)\\
  \nonum
  \dot y_i&=& x_i\;+\;a y_i\;+\;\epsilon(\bar y_i -y_i)\\
  \dot z_i&=&b+(x_i-c)z_i+\epsilon(\bar z_i-z_i)\;,
\eea
$i=1,...N$, as it was used in \cite{zanette2000}.

\subsection{Choice of order parameters}

As a first indicator for a partially or fully synchronized state, we measure the histogram of instantaneous pair distances
$d_{ij}(t)$ between all pairs of nodes as a function of (simulation) time, defined by \cite{zanette2000}
\bea
  \label{d}
  d_{ij}=\left[(x_i-x_j)^2 +(y_i-y_j)^2+(z_i-z_j)^2\right]^{1/2}\;,
\eea $i,j = 1,...N$. A fully synchronized state shows up as a
sharp peak in the distribution of $d_{ij}$, since the pair
distances between any two nodes approach zero. No synchronization
or desynchronization in the opposite case, lead to a broad
distribution. As order parameters in the usual sense (varying
between $0$ and $1$), $0$ for the desynchronized phase and $1$ for
the fully synchronized phase, we choose two order parameters $r$
and $s$, as proposed in \cite{zanette2000}, defined in the
following way. \bea
  \label{r}
  r(t)={1\over N(N-1)}\sum_{j\ne i,i,j=1}^N \Theta(\delta - d_{ij}(t)),
\eea
and
\bea
  \label{s}
  s(t)=1-{1\over N}\sum_{i=1}^N\prod_{j=1,j\ne i}^N \Theta(d_{ij}(t)-\delta),
\eea where $\Theta(x)$ is the Heavyside function, i.e.
$\Theta(x)=1$ if $x\ge 0$ and $\Theta(x)=0$ otherwise. The
parameter $\delta$ is a small number to account for the finite
numerical accuracy, e.g. $\delta = 0.0001$, so that two states in
phase space lying inside a sphere of radius $\delta$ are
considered as mutually being synchronized. The parameter $r(t)$
gives the fraction of pairs of elements $(i,j)$ which are
synchronized at time $t$ (i.e., $d_{ij}\leq \delta$). This
fraction is one if all possible pairs are synchronized and zero if
no pair is synchronized, intermediate values $0<r<1$ reflect
partial synchronization. The second order parameter $s(t)$ is more
sensitive to partial synchronization. The second term on the
r.h.s. of Eq.\ref{s} only contributes to the fraction if node $i$
has no other node within a distance of $\delta$. Therefore $s$ is
already $1$ when the total number of states is partitionized in
synchronized pairs without synchronization between the pairs. In
general we have $r<s<1$ (as it is confirmed in the figures below)
if some elements form clusters while others are still isolated.
From a simultaneous measurement of $r$ and $s$ it is possible to
obtain some information about the partial synchronization that is
usually a precursor to the fully synchronized state. We measured
in general all three functions $d_{ij}$, $r$, and $s$ as a
function of the number of iterations.

\section{Measurements and Results}

\subsection{Generating the topology}
For the scale-free topology we used the growth algorithm of B\'arabasi and Albert \cite{albert02},
later referred to as the BA model. In each step, one node with $m$
edges is added to the network. It is connected to $m$ of the formerly generated nodes according to preferential attachment.
In our simulations we chose $m$ between 1 and 10. For testing the role of the loops we used the regular topology of a
Cayley tree with $z$ edges at each node, $z=3,...,6$.The tree structure was then modified in various ways as we
shall see below. We also made some runs on a small-world topology, starting from a regular ring topology with $k=2$
neighbors and randomly adding shortcuts to each node with probability $p = 0.01$ according to the algorithm proposed
by Newman and Watts \cite{Newman99}.

\subsection{Choice of parameters}
For the parameters of the individual R\"ossler oscillators we chose $a=0.15$, $b=0.20$, $c=8.5$ throughout all simulations
to make sure that the individual systems are in the chaotic regime.The total number $N$ of oscillators was varied between
$10, 50, 200$ up to $500$ on the scale-free topology, and $N=190$ on the Cayley tree. The parameter $m$ of the
growth algorithm varied between $1$ and $10$, the coupling strength $\epsilon$ was out of the interval $[0.1,1.2]$.
In the numerical simulations of Eq.\ref{2} we used the fourth order Runge-Kutta methods with a typical time-step size
of $dt = 0.001$ (when $N=200$). Variation of $dt$ between $10^{-12}\leq dt \leq 10^{-1}$ led
to qualitatively the same results.

\subsection{Results for m=1}
Fig.1 a)
\begin{figure}
\includegraphics[width=8cm]{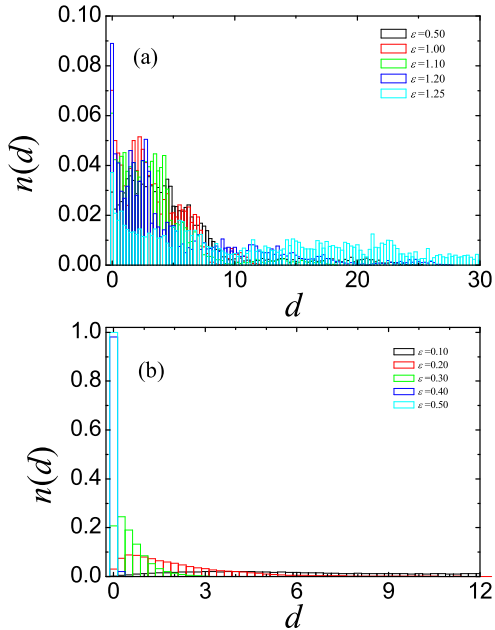}
\caption{Histogram of all pair distances between oscillators on a
BA network with (a) $m=1$, (b) $m=2$ for various $\epsilon$ and $N=200$}
\label{nd_BA_m_1}
\end{figure}
displays the results for the histogram of distances for $m=1$, $N=200$ and various couplings $\epsilon$ up to 1.25,
above which the numerical integration becomes unstable. The distributions are broad and do not indicate any synchronized state.
This result is further supported by measurements of $r$ and $s$ as shown in Fig. 2 a) and b), respectively,
\begin{figure}
\includegraphics[width=7.5cm]{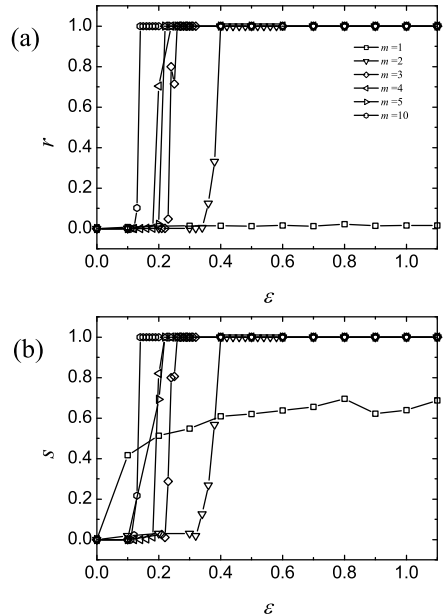}
\caption{Order parameters (a) $r$ and (b) $s$ as a function of the coupling strength $\epsilon$ for different values
of $m$, $N=200$}
\label{rs_BA_m_int}
\end{figure}
$r$ stays zero for $m=1$, while $s$ increases from $\epsilon = 0.1$ on, indicating some partial synchronization.
The value of $N=200$ seems to represent the large-$N$ limit, for the considered range of $\epsilon$ and $m$, since we obtained
the same result for $r$ and $s$ for $N = 300, 400, 500$. On the other hand, for smaller systems, $N<20$,
we do see a fully synchronized state when the
coupling $\epsilon$ exceeds a critical threshold. As value for $\delta$ in Eq.(\ref{r},\ref{s}) that accounts for the
finite numerical accuracy, we choose $\delta = 0.0001$. For too small values of $\delta$ we observe large variations
in the long-time behavior of $r$ and $s$, for too large $\delta$, the values of $r$ and $s$ are stable over long times,
but their values do depend on $\delta$. In between, i.e. for $0.0001<\delta<0.1$ we find a plateau for the values of $r$
and $s$, that is, the behavior becomes independent of the size of $\delta$.

\subsection{Results for m larger than one but still integer}
If we keep the number $N$ of oscillators fixed to $200$, we observe for $m>1$ a fully synchronized state above a
critical threshold in the coupling $\epsilon$; this threshold is the larger the smaller $m$, again $r<s$ in general,
as seen from Fig.2 a) and b).

\subsection{Results for intermediate noninteger m}
\begin{figure}
\includegraphics[width=7.5cm]{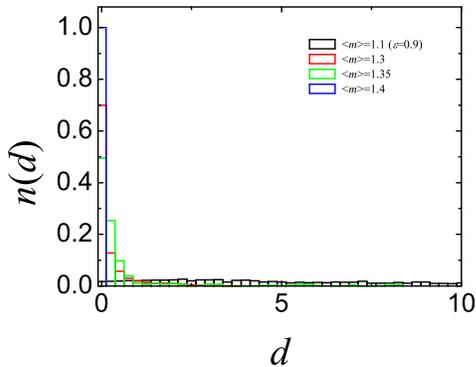}
\caption{Histogram of pair distances on a BA-model with
$1<\left<m\right><2$ with $N=200$ and $\epsilon=0.9$ }
\label{nd_BA_m_between1_2}
\end{figure}

\begin{figure}
\includegraphics[width=7.5cm]{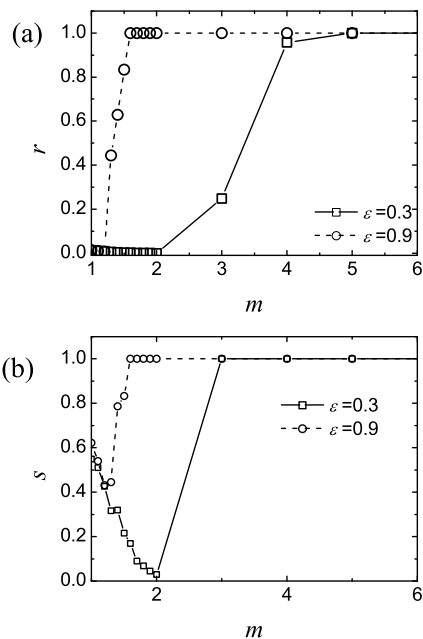}
\caption{Order parameters $r$ (a) and  $s$ (b) as function of the parameter $m$ that is used to distinguish different
topologies, for two values of the coupling strength}
\label{BA_rs_by_m}
\end{figure}

One of the main differences between the B\'arabasi-Albert networks with parameter $m=1$ and $m>1$ is the tree-like
structure for $m=1$ and the existence of loops for $m>1$. In order to check whether it is only the loops that facilitate
synchronization and how many loops are needed, we generalized the growth algorithm to non-integer values of $m$ in the
following way. We introduce an additional probability $p_m$ for a new node to have $m=1$ edges and $1-p_m$ for having
$m=2$ edges attached to the nodes of the network when it is introduced during the growth process. The distribution
of pair distances of oscillators for $1<\left< m\right><2$ is displayed in Fig.\ref{nd_BA_m_between1_2}.
For given $N$ and $\epsilon$ we therefore observe a transition to
a fully synchronized state as a function of "topology",
parameterized via the parameter $m$. For $N=200$ and $\epsilon
=0.9$, we have $1.35\leq m_c\leq 1.4$, cf.
Fig.\ref{nd_BA_m_between1_2}. Fig.\ref{BA_rs_by_m} a)
shows that the position of the transition, now in $m$ rather than
in $\epsilon$, depends on the coupling strength for fixed $N$. The
smaller $\epsilon$, the larger $m_c$. An interesting feature is
seen in Fig.\ref{BA_rs_by_m}b), where $s$ is plotted as a function
of $m$. For $\epsilon =0.3$ and $m=1$, the finite value of $s$
indicates some partial synchronization, $s$ then drops to zero at
$m=2$ and increases to $1$ for $m=3$. As we have argued above,
$s=1$ does not necessarily imply full synchronization, but some
partial one, at least. The behavior of $s$ is non-monotonic as a
function of $m$. A similar non-monotonic behavior of $s$ as
function of time was observed in \cite{mikhailovbook} for
R\"ossler oscillators, for which a partial synchronization was
followed by desynchronization, before the full synchronization set
in.

\subsection{R\"ossler oscillators on a Cayley tree}
\begin{figure}
\includegraphics[width=7cm]{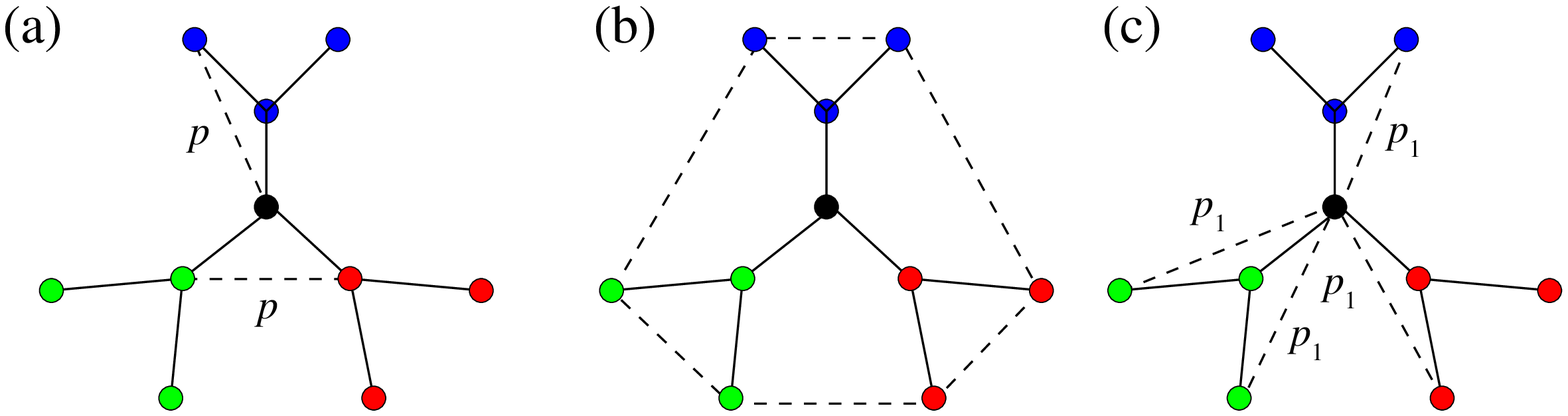}
\caption{Cayley tree for $z=3$ and additional interconnections (dashed lines) attached according to the three
rules a), b), c), respectively, as described in the text.} \label{bethe_ABC}
\end{figure}

\begin{figure}
\includegraphics[width=7cm]{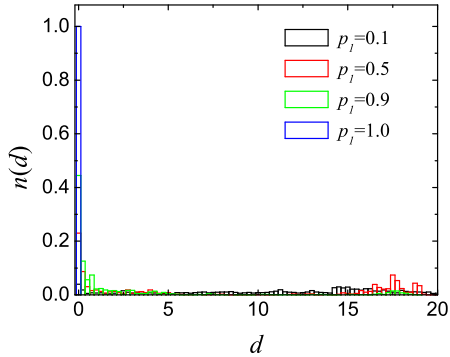}
\caption{Distance distributions for R\"ossler oscillators on a Cayley tree with additional edges according to c)
with different interconnection probabilities $p_1$, as further explained in the text} \label{Cayley_nd_p1}
\end{figure}

\begin{figure}
\includegraphics[width=7cm]{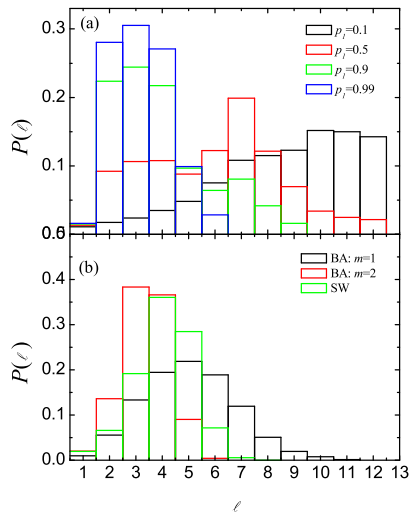}
\caption{Histograms of shortest paths of length $\ell$ for various networks with R\"ossler oscillators.
The envelope of the distributions looks similar for a) with $p=0.99$ and b) for $m=1, m=2$ and on the small world network} \label{short_path_dist}
\end{figure}

From the former results we conclude that a certain number of loops facilitates synchronization on scale-free networks, the
larger $m$, the more loops \cite{Bianconi03}, the smaller the coupling strength needed for synchronization.
In order to check whether it is the mere number of loops that facilitate synchronization or also the type of loops,
we studied
$N$ R\"ossler oscillators on a Cayley tree whose regular structure was modified in a controlled way by adding a) edges to
construct a given number of triangles at random locations, b) shortcuts between the outermost nodes, c) shortcuts between the
outermost nodes and the central node with probability $p_1$, see Fig.\ref{bethe_ABC}.
For $N\geq 187$ and arbitrary values of $\epsilon$ (more precisely, for $\epsilon$ as large as $\epsilon=0.9$),
R\"ossler oscillators do not synchronize
on a Cayley tree, neither for shortcuts according to a) or b), but for method c) and $p_1>0.9$ they do so, as it is seen
in Fig.\ref{Cayley_nd_p1}.
As it turns out also from measurements on a small-world topology with $p=0.01$, synchronization in all of our measurements
goes along with a distribution of shortest path lengths that looks Poissonian like, cf. Fig.\ref{short_path_dist}b)
On the other hand, it is not the {\it average} shortest path length that is conclusive alone.
As Fig. \ref{short_path_dist} shows, the average is almost the same in a) and b) while the distribution is different
and those of
Fig.\ref{short_path_dist}a) correspond to desynchronization on a Cayley tree without or with shortcuts between the
outermost nodes (apart from $p_1=0.99$),
while those of Fig.\ref{short_path_dist}b) lead to synchronization apart from the BA model with $m=1$, since in this case the
loops are totally absent. This shows that neither the existence of loops alone nor the existence of a certain
distribution of
shortest path lengths alone are sufficient to guarantee synchronization.

\section{Summary and Conclusions}
We studied an ensemble of R\"ossler oscillators on sale-free
networks constructed by the B\'arabasi-Albert growth algorithm. In
contrast to the usual investigations we studied the transition
from the desynchronized or partially synchronized state to the
fully synchronized state as a function of the network topology,
parameterized by the $m$, the number of newly attached edges in
the growth algorithm. For the tree topology ($m=1$) and given
coupling strength $\epsilon$, there is a fully synchronized state
below some critical size $N$ that disappears for larger $N$. This
result is similar to synchronization of Kuramoto oscillators on
Cayley trees which is possible for small enough size $N$ and
coordination number $z$ \cite{filippo}. Above a certain number of
nodes, the tree of R\"ossler oscillators is no longer
synchronizable however large the coupling strength is, but it is
then the parameter $m$ that introduces loops and shortcuts into
the tree, and along with this allows for full synchronization
again when $m$ exceeds a certain value that depends on $N$ and
$\epsilon$. The threshold in $\epsilon$ depends on $N$ and $m$,
and vice versa, the threshold in $m$ is sensitive to $N$ and
$\epsilon$. Small $N$, large $\epsilon$ (chosen from the stability
regime) and large $m$ favor synchronization. These qualitative
results are not specific for our choice of nonlinear couplings
between the R\"ossler oscillators, but also hold for the vector
coupling scheme of Eq.\ref{4}. Moreover, numerical simulations of
R\"ossler oscillators on Cayley trees with different artificially
introduced shortcuts suggest that it is not only the mere number
of loops that favors synchronization, but loops that provide real
shortcuts such as those between outermost and central nodes.

\end{document}